\documentclass[aps,prl,twocolumn,superscriptaddress,nofootinbib]{revtex4-2}
\usepackage{graphicx}
\usepackage[export]{adjustbox}
\usepackage{dcolumn}
\usepackage{xcolor} 
\usepackage{longtable}
\usepackage[caption=false]{subfig} 
\usepackage[colorlinks=true, linkcolor=black, citecolor=blue, urlcolor=black]{hyperref}
\bibpunct{\color{blue}[}{\color{blue}]}{,}{n}{}{,}

\begin{document}


\title{Uniaxial Compression-Induced Anisotropy and Electronic Dimensionality in the Iron-Based Superconductor FeSe}


\affiliation{Graduate School of Engineering, Kyushu Institute of Technology, Kitakyushu 804-8550, Japan}
\affiliation{Department of Applied Physics and Astronomy, University of Sharjah, U.A.E.} 
\affiliation{Lyman Laboratory of Physics, Harvard University, Cambridge, Massachusetts 02138, USA}

\author{Alexy Bertrand} 
\email{bertand.alexy-mathieu765@mail.kyutech.jp} 
\affiliation{Graduate School of Engineering, Kyushu Institute of Technology, Kitakyushu 804-8550, Japan}

\author{Masaki Mito}
\affiliation{Graduate School of Engineering, Kyushu Institute of Technology, Kitakyushu 804-8550, Japan}

\author{Kazuma Nakamura}
\affiliation{Graduate School of Engineering, Kyushu Institute of Technology, Kitakyushu 804-8550, Japan}
 
\author{Mahmoud Abdel-Hafiez}
\affiliation{Department of Applied Physics and Astronomy, University of Sharjah, U.A.E.} 
\affiliation{Lyman Laboratory of Physics, Harvard University, Cambridge, Massachusetts 02138, USA}

\begin{abstract}

The evolution of the superconducting transition temperature ($T_c$) in FeSe was investigated under in-plane, out-of-plane, and hydrostatic compression. For pressures up to 0.6 GPa, $T_c$ increases regardless of the compression mode, consistent with the suppression of nematic ordering. However, once nematicity is suppressed, $T_c$ exhibits a striking directional dependence: out-of-plane compression shows behavior similar to the hydrostatic case, with a sharp increase in $T_c$, whereas in-plane compression suppresses superconductivity. First-principles calculations suggest that in-plane compression shifts a hybridized band of Se $p_z$ and Fe $d_{x^2-y^2}$ character so that it crosses the Fermi level along the $\Gamma$-Z direction, leading to the emergence of an additional metallic band. This results in an increased three-dimensionality of the electronic structure and may be interpreted as a possible Lifshitz-type change in the Fermi surface. 

\end{abstract}


\maketitle

Iron-based superconductors (FeSCs) have attracted significant interest since the discovery of their high superconducting transition temperature ($T_c$) \cite{ref1}. In FeSCs, superconductivity usually occurs after antiferromagnetic (AFM) ordering is suppressed by chemical or pressure tuning \cite{ref2}. This AFM ordering appears in the parent compound at or below a temperature-driven phase transition from tetragonal to orthorhombic structure \cite{ref2,ref3}. Typical examples include the LaFeAsO family \cite{ref4,ref5}, BaFe2As2 family \cite{ref6,ref7}, and the NaFeAs family \cite{ref8}. 

As the parent compound of the Fe-chalcogenide superconductors (also known as the 11-system), FeSe has a relatively simple crystal structure and exhibits a superconducting transition temperature $T_c$ of 9 K \cite{ref9}. However, no AFM ordering is observed below the tetragonal-to-orthorhombic structural transition at around 90 K \cite{ref10,ref11}. This phase, with no magnetic ordering but in-plane anisotropy, is referred to as the nematic phase. The electronic anisotropy of the nematic phase has been observed using resistivity measurements \cite{ref12}, optical measurements \cite{ref13}, or magnetic susceptibility measurements \cite{ref14}.

The behavior of $T_c$ in FeSe under pressure has been studied by many groups using electrical conductivity or magnetization measurements \cite{ref15,ref16,ref17,ref18,ref19,ref20}. FeSe exhibits a characteristic three-step evolution of $T_c$ under hydrostatic pressure. Initially, $T_c$ rises to a local maximum of 13 K at 0.8 GPa, due to the gradual suppression of nematicity \cite{ref18,ref19,ref20}. At 0.8 GPa, an AFM ordering appears, and then $T_c$ decreases until reaching a local minimum of about 10 K at 1.2 GPa \cite{ref16,ref19,ref21}. As pressure increases further, magnetic ordering diminishes, and $T_c$ rises sharply until reaching a maximum of 38.3 K at 6 GPa, when magnetic ordering is entirely suppressed \cite{ref19}. Beyond this, $T_c$ drops again, due to FeSe's tendency to transform into a new orthorhombic structure, ultimately leading to the loss of superconductivity \cite{ref22,ref23}. In FeSe, there is a correlation between the change in $T_c$ and the height of the Se atom $h = c \times \text{Se}(z)$ \cite{ref24}. Under hydrostatic pressure, the decrease in $h$ is associated with an increase in $T_c$. This distinctive behavior underscores the complex interplay among nematicity, magnetic order, and superconductivity in FeSe.

Another key feature of FeSe is the significant increase in $T_c$ to above 100 K in monolayer thin films \cite{ref25,ref26}. $T_c$ is highly sensitive to the choice of substrate and drops rapidly with increasing thickness \cite{ref27}. Therefore, not only hydrostatic pressure but also uniaxial stress, concerning the specimen geometry, are essential factors for tuning the superconductivity of FeSe. Other researchers have explored the effects of in-plane tensile and compressive strains, accompanied by the Poisson effect, on FeSe thin films \cite{ref28,ref29,ref30} and single crystals \cite{ref31,ref32}. 

However, a thorough experimental study comparing the effects of hydrostatic and uniaxial compression, without the accompanying Poisson effect, across different crystal directions remains missing. Obtaining purely uniaxial data is essential for a comprehensive understanding of the mechanisms governing the three-step pressure evolution of $T_c$. This study examined the effect of uniaxial strain on FeSe and compared it with the effect of hydrostatic compression. The focus was on the pressure range near the disappearance of the nematic phase and the emergence of the AFM phase.

The magnetization of FeSe single crystals under different compression modes was measured inside a miniature diamond anvil cell (mDAC) made of a CuBe alloy and designed for insertion into a commercial superconducting quantum interference device (SQUID) magnetometer \cite{ref33}. The top diameter of the diamond anvils was 0.8 mm, and a hardened CuBe gasket with a sample space of 0.4 mm was used. A liquid-like pressure-transmitting medium, Daphne oil 7373, was used to realize the hydrostatic pressure conditions. To achieve the uniaxial compression, epoxy resin (Stycast 1266) was mixed with a precursor and inserted into the sample space \cite{ref34}. It was then placed inside a drying chamber at 30$^\circ$C for about 1 day, until complete solidification. 'Out-of-plane' compression was applied by setting a single-crystal sheet on the gasket plane. 'In-plane' compression was applied by placing multiple sheets perpendicular to the gasket plane. Indeed, the orientations of $a$ and $b$ within the mDAC cannot be determined. Thus, the direction of in-plane compression is $a$, $b$, or a combination of both. The [001] orientation of a FeSe sheet was checked using x-ray diffraction and is reported in supplementary material \textcolor{red}{Fig. S1}. Images of the setting inside the gasket hole for both uniaxial compression experiments are shown in \textcolor{red}{Fig. 1}. 

A SQUID magnetometer equipped with an ac option was used to measure the ac magnetization $m$ between $T = 5$ and 40 K using an ac field at 10 Hz and an amplitude of 3.9 Oe.  $T_c$ was investigated by examining the magnetic shielding signature in the ac magnetization $m$, which is divided into an in-phase magnetization $m'$ and an out-of-phase magnetization $m''$ by Fourier transformation. The signal due to the metallic background of the mDAC mainly appears in $m''$ and the magnetic shielding signal of the samples in $m'$ \cite{ref35,ref36}. The remaining metallic background in $m'$ can be subtracted using the signal obtained without any sample inserted.

\begin{figure}
\includegraphics[width=\columnwidth]{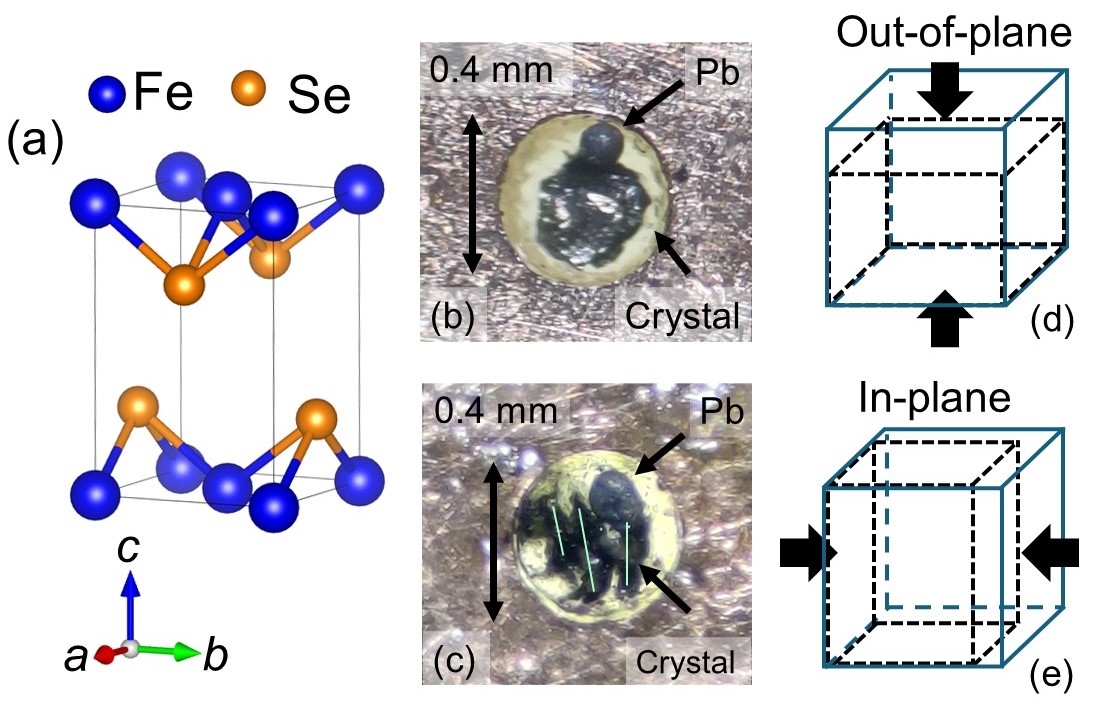}
\caption{\label{fig1}(a) Structure of tetragonal FeSe. Picture of the gasket hole showing the setting of the single crystals for the (b) in-plane and (c) out-of-plane compression measurements, and (d-e) the respective schematics of their impact on the unit cell.}
\end{figure}


The applied pressure or stress was estimated from the pressure dependence of lead's superconducting temperature \cite{ref37}. Indeed, we recognize that pressure should only be defined in the cases of hydrostatic compression.  However, the amount of strain cannot be measured inside the mDAC. Only the magnitude of stress can be controlled. Thus, pressure was used as the comparative experimental parameter for hydrostatic and uniaxial compression conditions, as reported in the literature \cite{ref34,ref35,ref38,ref39}.

To discuss the observed $T_c$ trend, the ab initio density functional calculations using the Quantum Espresso package \cite{ref40,ref41} were performed to determine the band structure under the different strain conditions. Norm-conserving Pseudo potentials obtained from Pseudo Dojo \cite{ref42,ref43} were employed together with the generalized gradient approximation for the exchange-correlation potential \cite{ref44}. The cutoff energies for the wavefunctions and charge densities were 100 Ry and 400 Ry, respectively. Brillouin zone integration was carried out using a $8 \times 8 \times 8$ k-mesh. The Fermi energy was evaluated using Gaussian smearing with a smearing width of 0.136 eV. For simplicity, we considered the tetragonal structure and neglected the orthorhombic distortion, which does not affect the overall trend. The analysis based on maximally localized Wannier functions was performed using the RESPACK code \cite{ref45}.

The pressure dependence of the magnetization $m'$ of FeSe for the different compression modes is shown in  \textcolor{red}{Fig. 2}. At 0 GPa, the measured $T_c$ is 9.8 K, which is slightly higher than the 9 K found in the literature for magnetization measurements \cite{ref18,ref19}. This difference can be explained by the method used to determine $T_c$. Moreover, FeSe may already be under a pressure of around 0.1 GPa because the crystal was cooled while surrounded by Daphne 7373 inside the mDAC, thereby slightly increasing $T_c$. Indeed, FeSe shows high ductility at pressures below 1 GPa \cite{ref46}.

\begin{figure*}
\includegraphics[width=0.95\textwidth]{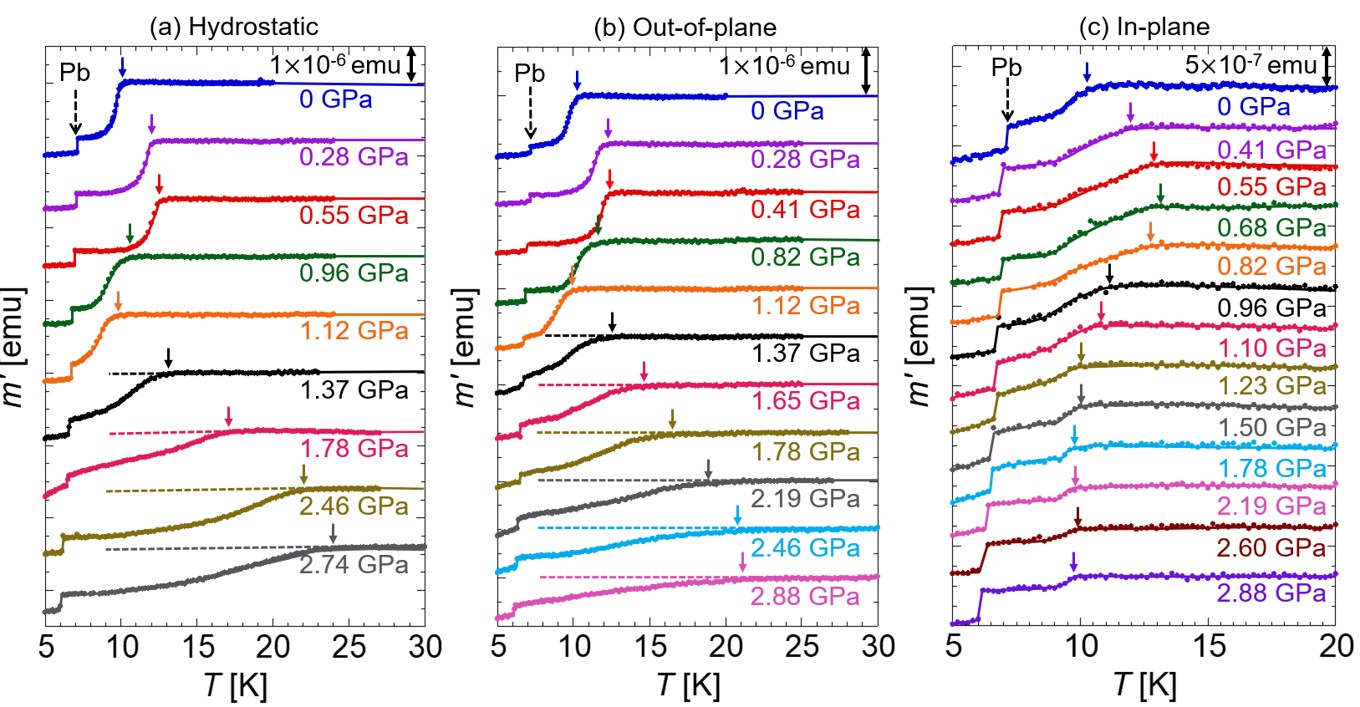}
\caption{\label{fig2}Temperature dependence of the in-phase magnetization $m'$ of FeSe for pressure up to 3 GPa using Pb as a manometer in the cases of (a) hydrostatic pressure conditions, (b) out-of-plane uniaxial compression, and (c) in-plane uniaxial compression. The small colored arrows indicate the temperature at which the magnetic shielding anomaly of FeSe appears. The magnetic signal from the cell without samples was subtracted, resulting in flat curves over $T_c$. The sharp signal at low temperature, indicated by a black dashed arrow, is the magnetic shielding signal of Pb. Pressure calibration details are shown in \textcolor{red}{Fig. S2}.}
\end{figure*}

As shown in \textcolor{red}{Fig. 2(a)}, a three-step-like change in $T_c$ of FeSe under pressure is observed, consistent with measurements from other groups \cite{ref18,ref19,ref20}. A first maximum is present at around 0.5 GPa at 12 K and corresponds to the balance between the disappearance of the nematic phase and the appearance of the AFM ordering. Then, there is a minimum at around 1.1 GPa of 10.2 K, corresponding to the point at which the AFM ordering begins to decrease. Finally, $T_c$ increases sharply up to 2.74 GPa, reaching 24 K. A sudden broadening of the superconducting signal is observed above 1 GPa and cannot be explained solely by the pressure distribution within the mDAC setup. This is inherent to FeSe and has been suggested to arise from the competition between the emergence of the AFM ordering and superconductivity \cite{ref18}.

The pressure dependence of $T_c$ shows a similar trend under out-of-plane uniaxial compression. As shown in \textcolor{red}{Fig. 2(b)}, at pressures below 1 GPa, $T_c$ values are nearly identical to those in the hydrostatic case. At higher pressure, a significant increase in $T_c$ is also observed. The maximum increase is smaller, reaching 21 K at 2.88 GPa.  Given the method used to associate pressure with uniaxial compression, the difference in $T_c$ between out-of-plane and hydrostatic compression is not considered significant.

\textcolor{red}{Figure 2(c)} shows the dependence of $m'$ on the estimated pressure for the in-plane compression mode. The smaller observed superconducting signal is due to the reduced sample volume in the gasket hole and to the orientation of the FeSe sheets, which affects the demagnetization factor. For pressures below 1 GPa, the same maximum in $T_c$ occurs around 0.68 GPa. These results indicate that, similar to other compression modes, in-plane compression suppresses the nematic phase. In-plane compressive measurements of FeSe, including Poisson effect, and for strain up to 1.5\% in thin films and 0.1\% in bulk samples, have shown an increase in $T_c$ \cite{ref31}. This change in $T_c$ is similar to that observed at pressures below 0.68 GPa in our study. From 1.1 GPa onward, the behavior differs significantly from that in other compression modes, with $T_c$ decreasing with increasing pressure, reaching 9.5 K at 2.88 GPa. At the same time, the superconducting signal intensity also decreases, suggesting possible instability in the superconducting phase.
\begin{figure}
\includegraphics[width=0.85\columnwidth]{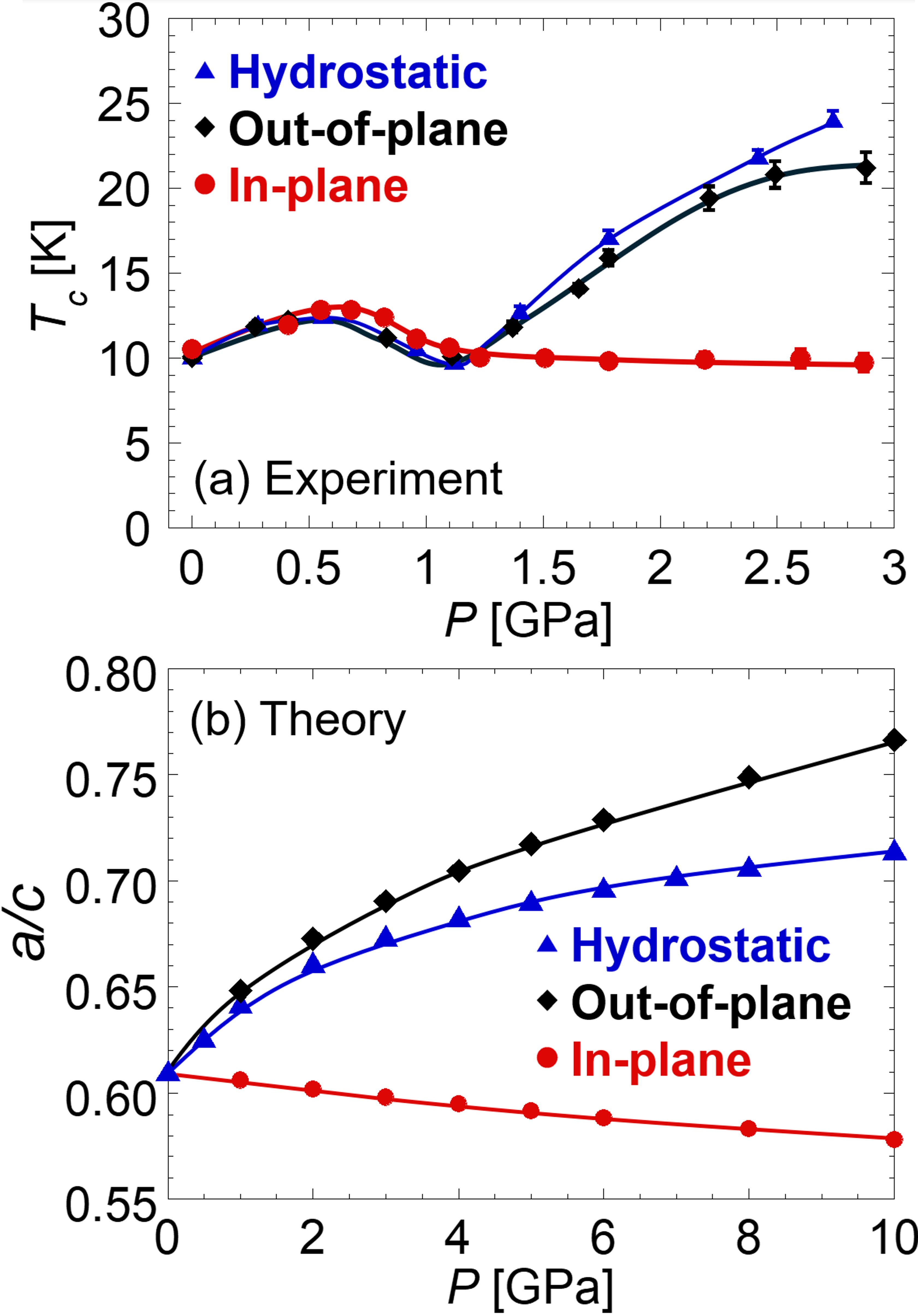}
\caption{\label{fig3}(a) Experimental pressure dependence of the superconducting temperature $T_c$ obtained from \textcolor{red}{Fig. 2} and (b) theoretical pressure dependence of the tetragonal distortion $a/c$ of FeSe for the different compression modes. Blue triangles correspond to the hydrostatic pressure case, black diamonds to the out-of-plane uniaxial compression, and red circles to the in-plane uniaxial compression.}
\end{figure}

The $T_c$ dependence for each compression mode is summarized in \textcolor{red}{Fig. 3(a)}. Below approximately 1 GPa, no significant difference is observed between the compression modes. Above this pressure, however, clear differences emerge: $T_c$ increases under hydrostatic and out-of-plane compression, whereas it remains nearly constant under in-plane compression. The behavior below 1 GPa may be related to the nematic state of FeSe. In contrast, above about 1 GPa, AFM ordering has been reported under hydrostatic compression \cite{ref21}, and the emergence of magnetic correlations may be reflected in the observed differences between the compression modes.

\textcolor{red}{Figure 3(b)} shows the pressure dependence of the tetragonal distortion $a/c$ obtained from first-principles calculations. To model the out-of-plane uniaxial strain, the $a$ parameter obtained from the ambient-pressure case was fixed during optimization.  Similarly, $c$ was fixed for in-plane strain. Previous hydrostatic compression experiments showed that $a/c$ correlates with $T_c$ \cite{ref23}. Here also, the behavior of $a/c$ correlates well with the $T_c$ dependence observed above 1 GPa in \textcolor{red}{Fig. 3(a)}. This parameter can be regarded as an indicator of the structural dimensionality. A larger value of $a/c$ corresponds to a more three-dimensional structure, whereas a smaller value indicates a more two-dimensional character. This suggests that the structural dimensionality characterized by $a/c$ plays an important role in determining the superconducting behavior in the high-pressure regime. While $a/c$ predicts a consistent trend for $T_c$, the electronic band structure will ultimately govern the superconducting state.

\begin{figure}
\includegraphics[width=0.987\columnwidth, right]{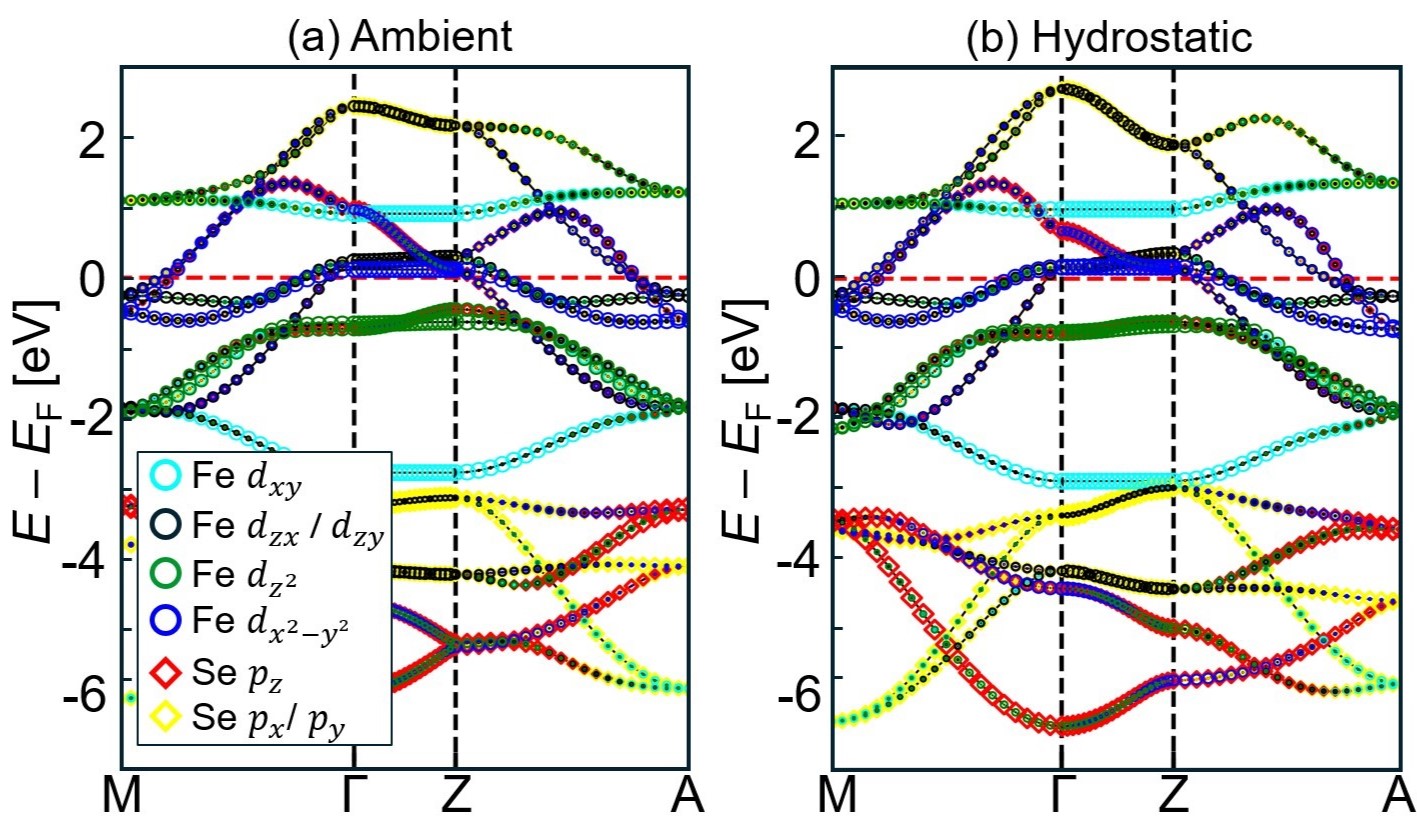}
\includegraphics[width=0.982\columnwidth, right]{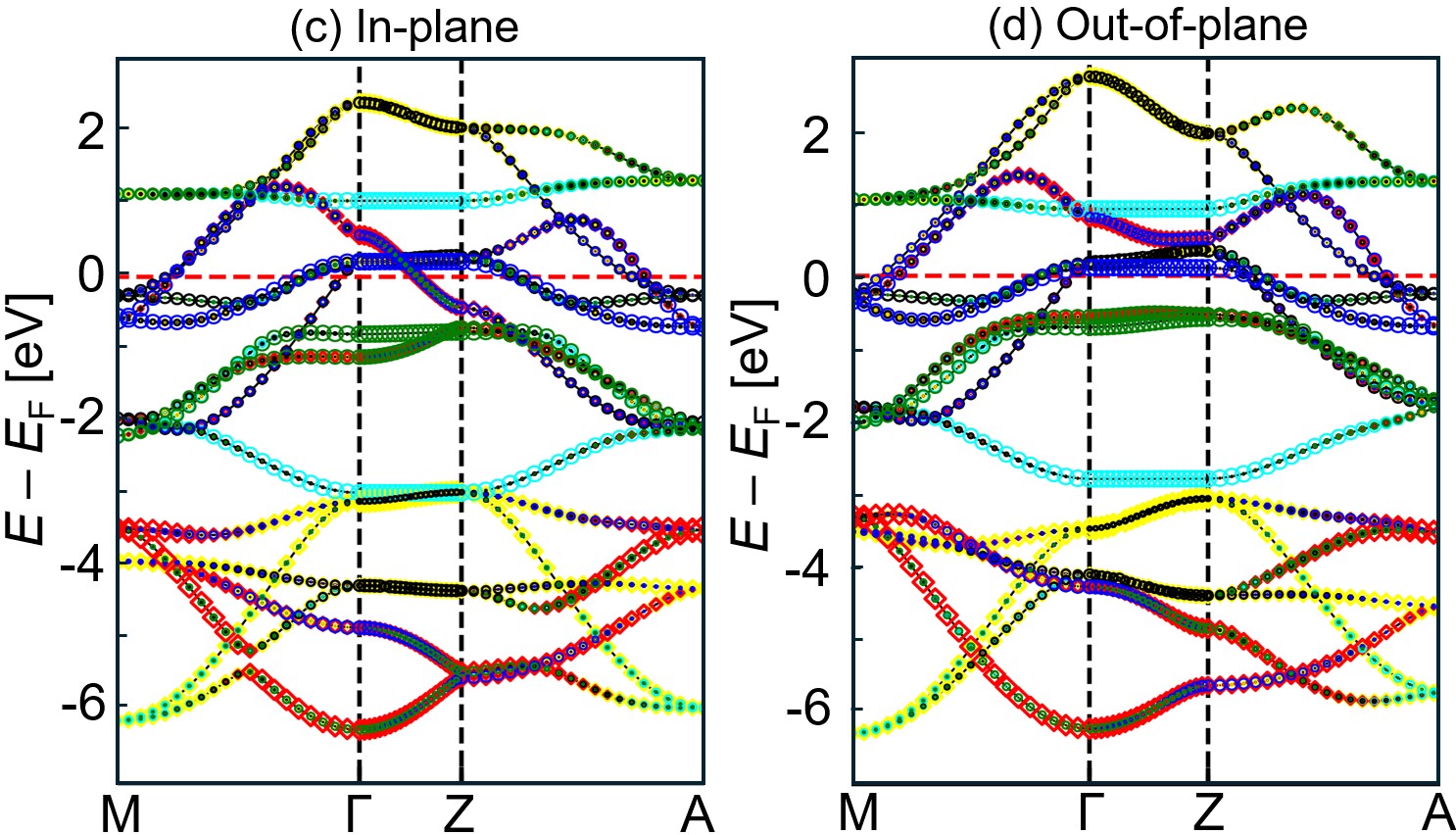}
\includegraphics[width=\columnwidth]{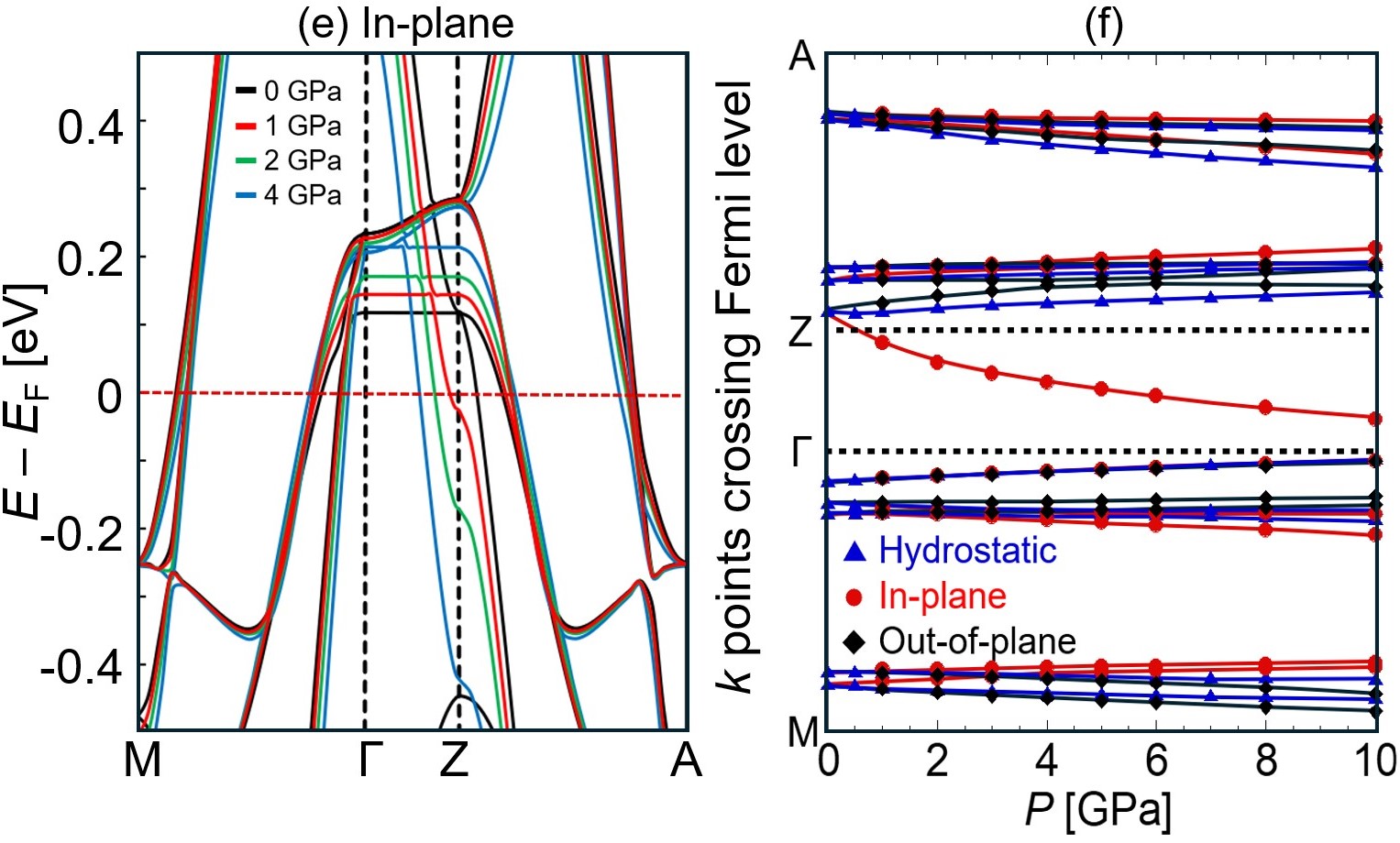}
\caption{\label{fig4}Fat-band representations based on maximally localized Wannier functions for FeSe under (a) ambient pressure, (b) hydrostatic pressure, (c) in-plane uniaxial compression, and (d) out-of-plane uniaxial compression with an applied pressure of 4 GPa. The Fermi energy is set to zero. (e) Enlarged view of the band structure near the Fermi level under in-plane uniaxial compression as a function of pressure. (f) Pressure dependence of the k points where the bands cross the Fermi level for the three compression modes along the M–$\Gamma$–Z–A path. Detailed calculations are shown in \textcolor{red}{Figs. S3-S9}.}
\end{figure}

Here, to investigate the pressure-mode dependence of the electronic structure, we performed a fat-band analysis based on maximally localized Wannier functions \cite{ref45}. \textcolor{red}{Figure 4(a–d)} shows the orbitally decomposed band structures for each compression mode. Although the overall features of the band structures are very similar, a notable difference appears along the $\Gamma$-Z direction near the Fermi level. In the case of the in-plane compression, shown in \textcolor{red}{Fig. 4(c)}, a band crosses the Fermi level near the middle of the $\Gamma$-Z path, whereas no such crossing is observed under the other compression modes. This band is found to have a hybridized character of Se $p_z$ and Fe $d_{x^2-y^2}$ orbitals, as shown in \textcolor{red}{Figure 4(a–d)}. This indicates that the electronic structure under in-plane compression has a more three-dimensional character.  In contrast, the other compression modes exhibit more two-dimensional electronic structures with flat bands along the $\Gamma$-Z path. A significant discrepancy emerges between structural and electronic character. While in-plane compression results in a structurally more two-dimensional unit cell (smaller $a/c$), fat-band analysis confirms the emergence of a new metallic band along the $\Gamma$-Z direction, rendering the electronic structure more three-dimensional.

\textcolor{red}{Figure 4(e)} shows the pressure dependence of the electronic structure under in-plane compression. As seen in the figure, the degeneracy at the Z point is lifted as pressure increases. This splitting causes one of the bands to cross the Fermi level along the $\Gamma$–Z path. \textcolor{red}{Figure 4(f)} shows the pressure dependence of the k-points at which the Fermi level is crossed along the band diagram. For hydrostatic and out-of-plane compression, no Fermi-level crossing is observed along the $\Gamma$-Z path. In contrast, a new crossing appears along this path under in-plane compression, highlighting a topological change in the Fermi surface. 

The compression-mode dependence of the low-energy electronic structure can be discussed as follows. The suppression of nematic ordering under pressure may enhance magnetic fluctuations, which could in turn promote superconductivity. The additional metallic channel appearing under in-plane compression likely triggers a redistribution of orbital occupancy. This electronic reconstruction prevents the expected $T_c$ enhancement seen in hydrostatic conditions, suggesting that the preservation of electronic two-dimensionality is vital for high-$T_c$ superconductivity in Fe-chalcogenides. This behavior may also be interpreted as a possible Lifshitz-type reconstruction of the Fermi surface associated with the emergence of the additional metallic band.

Replacing Se atoms with smaller S atoms to produce FeSe$_{1-x}$S$_x$ results in the suppression of the nematic state. Moreover, the magnetic ordering appears at higher pressure as x increases \cite{ref24,ref48,ref49,ref50,ref51}. In FeSe$_{0.9}$S$_{0.1}$, AFM ordering occurs only above 3 GPa, and a pressure-$T_c$ dependence similar to FeSe is observed. This indicates that the increase in $T_c$ at pressures above 1 GPa may not be correlated with the suppression of the magnetic ordering. Indeed, our results on FeSe suggest that the $T_c$ enhancement above 1 GPa under hydrostatic and out-of-plane conditions is an intrinsic structural effect, whereas the suppression of $T_c$ under in-plane strain is an electronic effect. Similarly, Te doping in FeSe$_{1-x}$Te$_x$ also controls the superconducting properties \cite{ref52,ref53,ref54}. Thus, comparing uniaxial compression experiments for the present FeSe case with those for FeSe$_{0.9}$S$_{0.1}$ and FeSe$_{0.9}$Te$_{0.1}$ would provide crucial information on the mechanism underlying the pressure-induced increase in $T_c$. 

This study uncovers an anisotropic dependence of $T_c$ on uniaxial compression, potentially linked to structural and electronic dimensionality. This highlights the importance of uniaxial compression experiments for investigating the mechanism of superconductivity in FeSe. Since FeSe is widely regarded as a strongly correlated iron-based superconductor \cite{ref55}, theoretical approaches that explicitly incorporate electron correlation effects are essential for understanding its nematic phase and superconductivity. Uniaxial strain effects, including the Poisson effect, have also been considered in calculations based on dynamical mean-field theory \cite{ref56}. Our results suggest that, for realistic modeling of FeSe under strain, it may also be necessary to explicitly consider the degrees of freedom associated with the Se $p$ orbitals. The hybridized Se $p_z$ and Fe $d_{x^2-y^2}$ band exhibits strong dispersion along the $c$ axis, and its shift under in-plane compression plays an important role in the emergence of the additional metallic band.

\bigskip

\begin{acknowledgments}
This study was supported by a Grant-in-Aid for Scientific Research (Grant No. 23H01126) from MEXT, Japan. The data that support the findings of this study are available from the corresponding author upon reasonable request. 
\end{acknowledgments}

\bibliography{reference.bib}

\clearpage
\widetext
\begin{center}
\textbf{\LARGE Supporting information}
\end{center}

\setcounter{figure}{0}
\setcounter{table}{0}
\setcounter{page}{1}
\makeatletter

\renewcommand{\thefigure}{S\arabic{figure}}
\renewcommand{\thetable}{S\arabic{table}}
\renewcommand{\citenumfont}[1]{S#1}

\begin{figure}[h]
\includegraphics[width=0.5\textwidth]{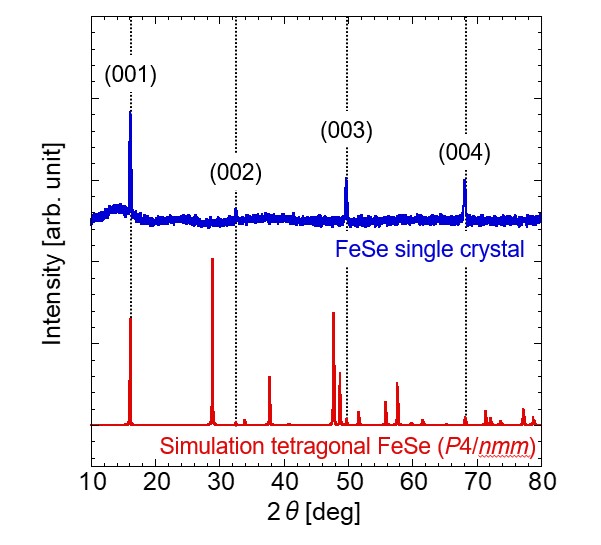}
\caption{\label{figS1} X-ray diffraction patterns of the FeSe single crystals used for the magnetization measurements as set in the gasket hole of the diamond anvil cell. A simulation pattern for the tetragonal $P4/nmm$ phase of FeSe is represented in red. Only the (00$l$) peaks are visible, confirming the orientation of the single crystals.}
\end{figure}

\begin{figure}[h]
\includegraphics[width=0.66\textwidth]{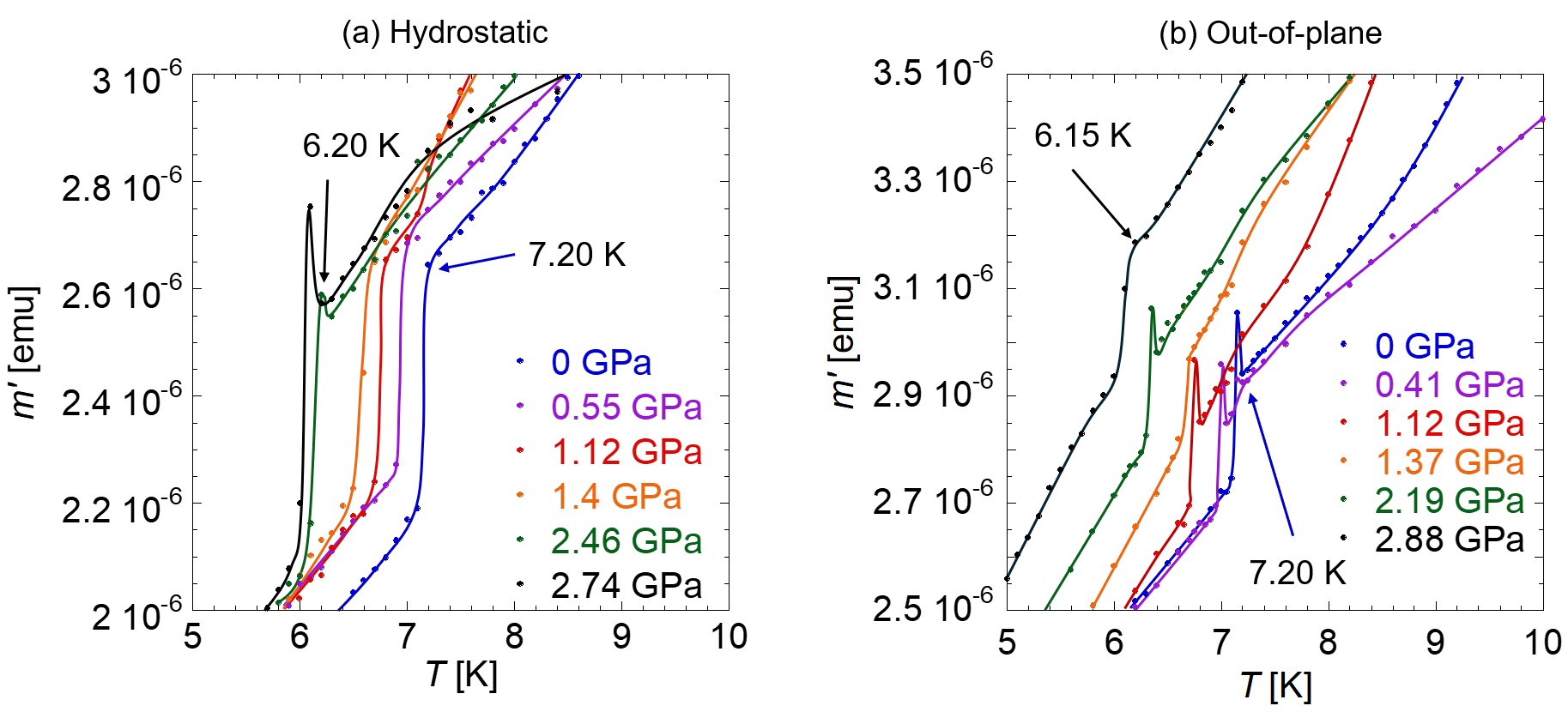}
\includegraphics[width=0.32\textwidth]{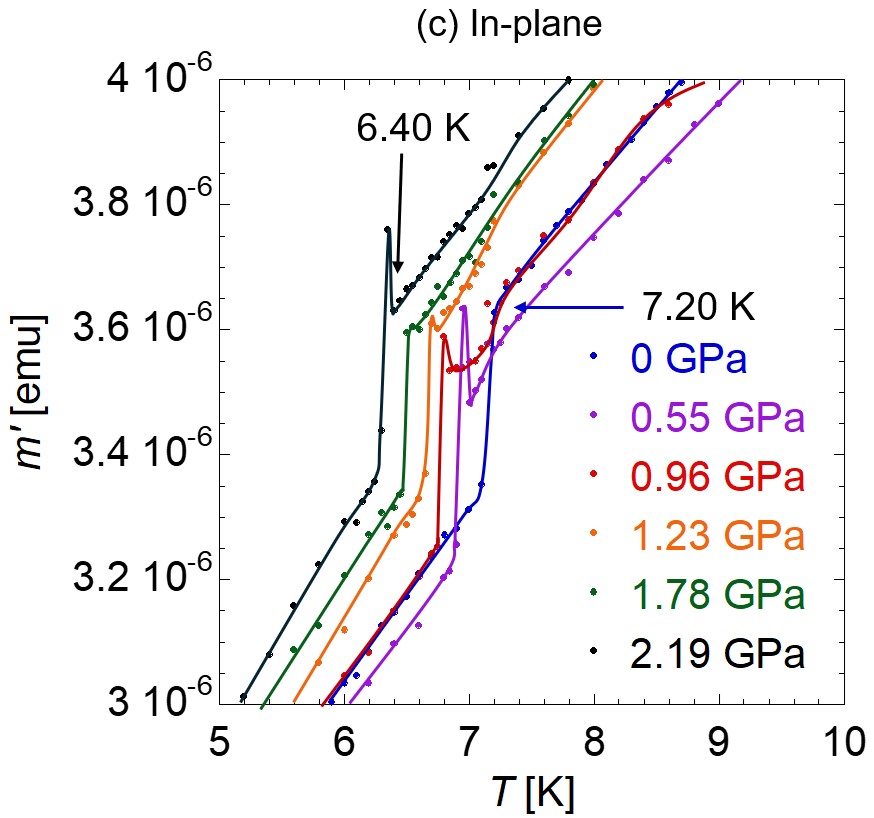}
\caption{\label{figS2}Magnified view of the temperature dependence of the in-phase magnetization $m’$ of FeSe in the cases of (a) hydrostatic pressure conditions, (b) out-of-plane uniaxial compression, and (c) in-plane uniaxial compression. m’ before background subtraction is shown. The sharp signal at 7.2 K at ambient pressure is the magnetic shielding signal of Pb. Here, only six measurements are shown in all cases to illustrate the shift of the $T_c$ of Pb toward the low-temperature side. The pressure is then estimated using the formula: $P$ [GPa] = (7.2 – $T_c$ [K])/0.365. (M. J. Clark and T. F. Smith, J. Low Temp. Phys. 32, 495 (1978)). Before the magnetic shielding anomaly, a sharp increase in m’ is sometimes observed and inherent to the Pb signal. For consistency, $T_c$ of Pb is defined as the last data point before this anomaly.}
\end{figure}

\begin{table*}[t]
\renewcommand{\arraystretch}{1.3}
\caption{\label{tab:lattice_params} Lattice parameters and atomic positions Se(z) of tetragonal FeSe under hydrostatic, out-of-plane, and in-plane compressions obtained after structural optimization and used for band calculations.}
\begin{ruledtabular}
\begin{tabular}{cccccccc}
 & \multicolumn{3}{c}{Hydrostatic} & \multicolumn{2}{c}{Out-of-plane ($a =  3.6761$~\AA)} & \multicolumn{2}{c}{In-plane ($c = 6.0209$~\AA)} \\
 \cline{2-4} \cline{5-6} \cline{7-8}
 Pressure [GPa] & $a$ [\AA] & $c$ [\AA] & Se($z$) & $c$ [\AA] & Se($z$) & $a$ [\AA] & Se($z$) \\
 \hline
 0  & 3.6761 & 6.0209 & 0.2300 & -- & -- & -- & -- \\
 1  & 3.6589 & 5.6950 & 0.2442 & 5.6713 & 0.2436 & 3.6497 & 0.2324 \\
 2  & 3.6463 & 5.5109 & 0.2528 & 5.4644 & 0.2520 & 3.6238 & 0.2349 \\
 3  & 3.6356 & 5.3919 & 0.2587 & 5.3253 & 0.2577 & 3.6008 & 0.2370 \\
 4  & 3.6252 & 5.3046 & 0.2632 & 5.2161 & 0.2621 & 3.5807 & 0.2387 \\
 5  & 3.6157 & 5.2346 & 0.2669 & 5.1245 & 0.2658 & 3.5610 & 0.2404 \\
 6  & 3.6071 & 5.1741 & 0.2702 & 5.0439 & 0.2691 & 3.5425 & 0.2420 \\
 8  & 3.5905 & 5.0782 & 0.2755 & 4.9094 & 0.2744 & 3.5098 & 0.2447 \\
 10 & 3.5753 & 5.0014 & 0.2799 & 4.7965 & 0.2789 & 3.4798 & 0.2471 \\
\end{tabular}
\end{ruledtabular}
\end{table*}

\begin{figure}[h]
\includegraphics[width=0.7\textwidth]{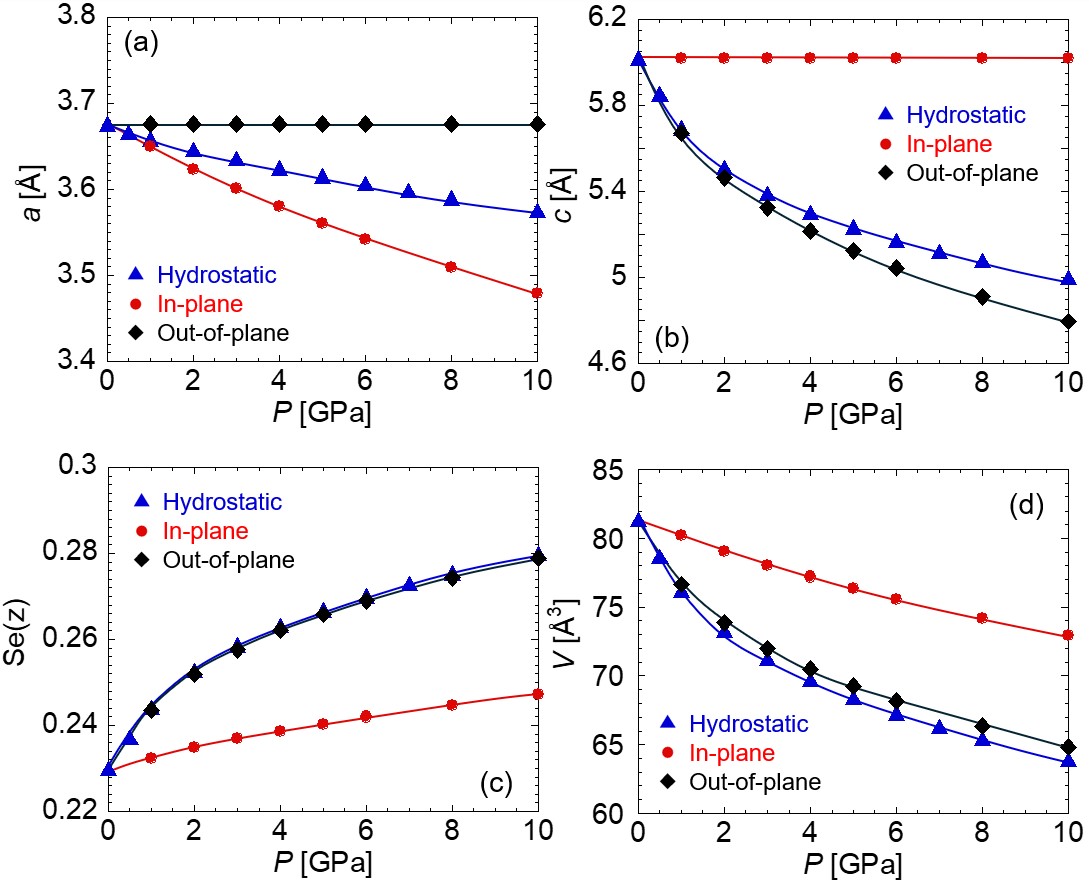}
\caption{\label{figS3}(a) (b) Lattice parameters, (c) atomic position Se(z), and (d) unit cell volume depending on pressure and compression type plotted using the data in Table S1.}
\end{figure}

\begin{figure}[h]
\includegraphics[width=\textwidth]{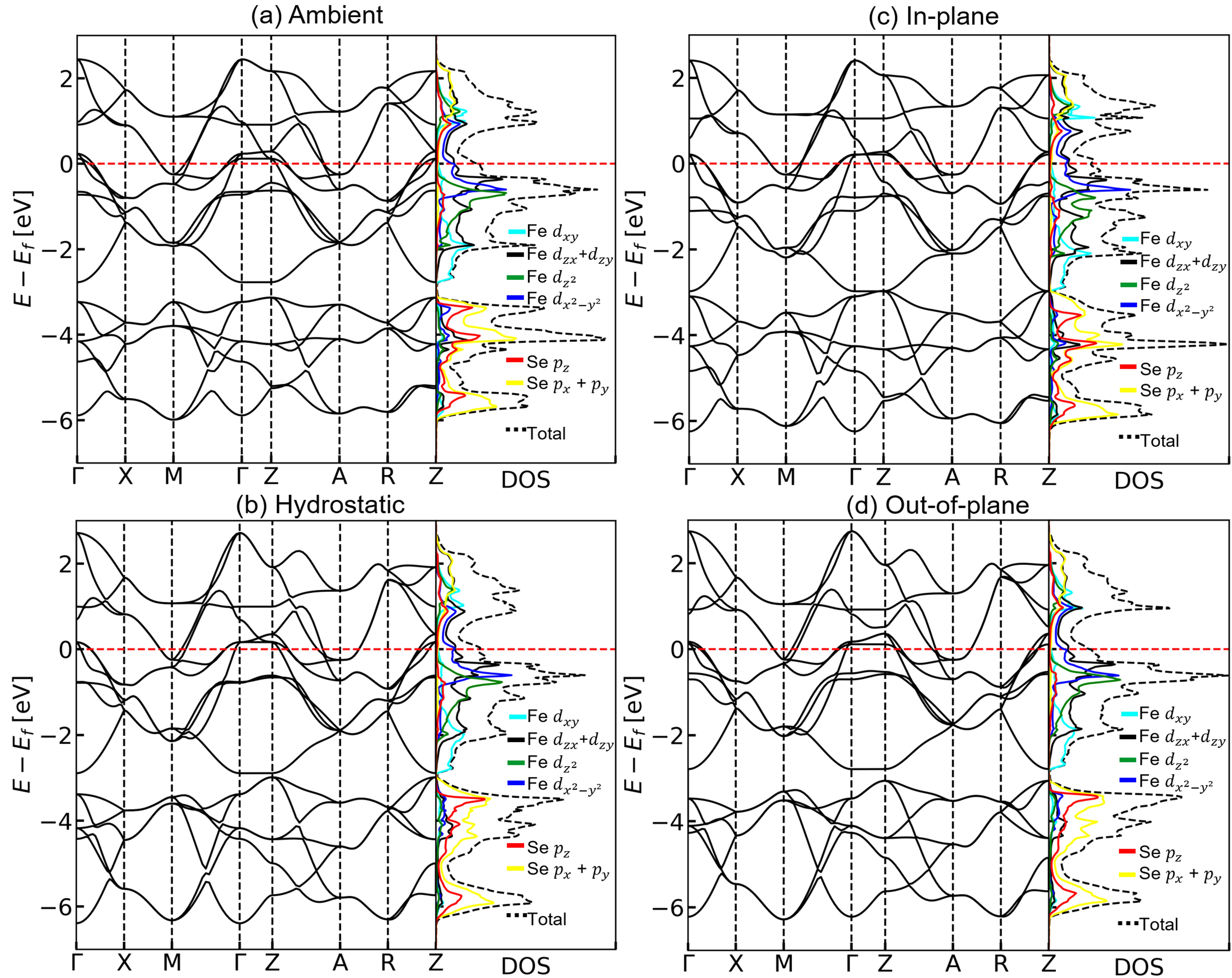}
\includegraphics[width=0.4\textwidth]{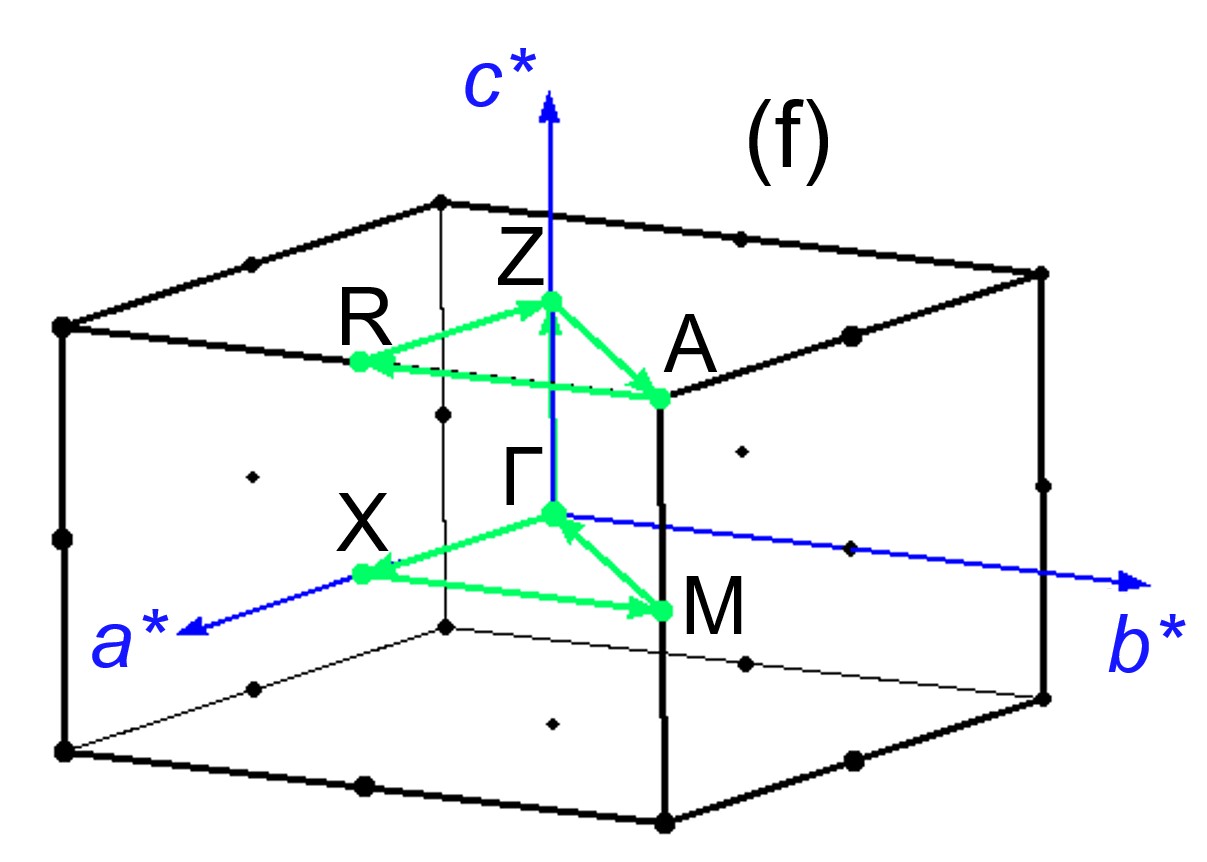}
\caption{\label{figS4}Density-functional band structure and calculated electronic density of states for FeSe under (a) ambient pressure, (b) 4 GPa of hydrostatic pressure, (c) out-of-plane uniaxial compression, and (d) in-plane uniaxial compression calculated using maximally localized Wannier functions. Each graph is aligned to the Fermi energy, indicated by a dotted red line. For the density of states, the electron population is obtained by integrating over the considered energy scale. The total electron population in black is obtained by summing the populations of Fe (blue) and Se (orange). The Brillouin zone used for calculation is shown in (e).}
\end{figure}

\begin{figure}[h]
\includegraphics[width=\textwidth]{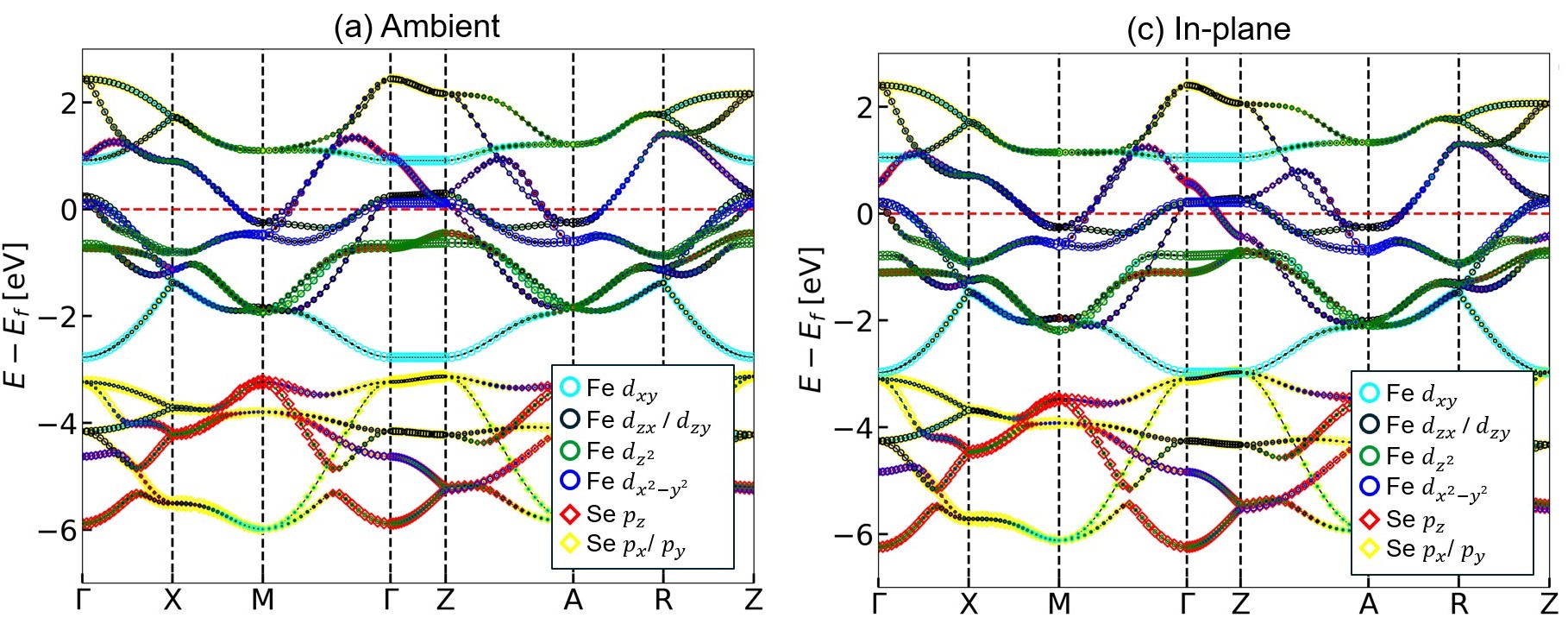}
\includegraphics[width=\textwidth]{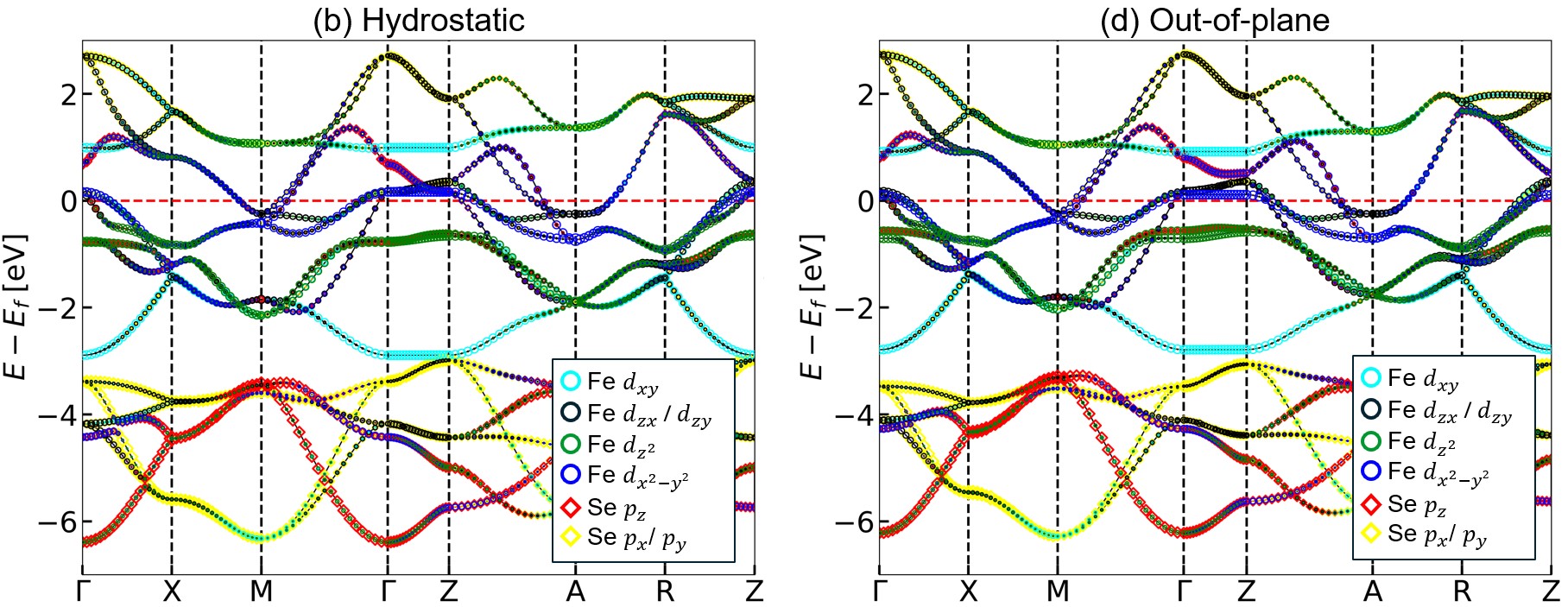}
\setkeys{Gin}{width=0.48\linewidth}
    \subfloat{\includegraphics{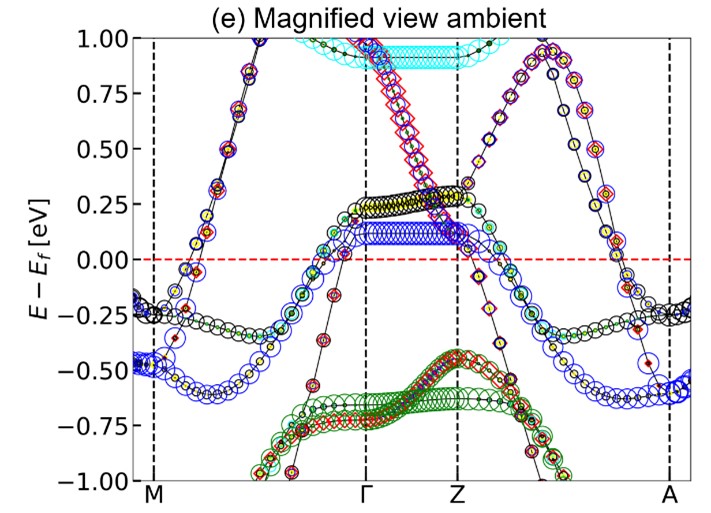}}\hfill%
    \subfloat{\includegraphics{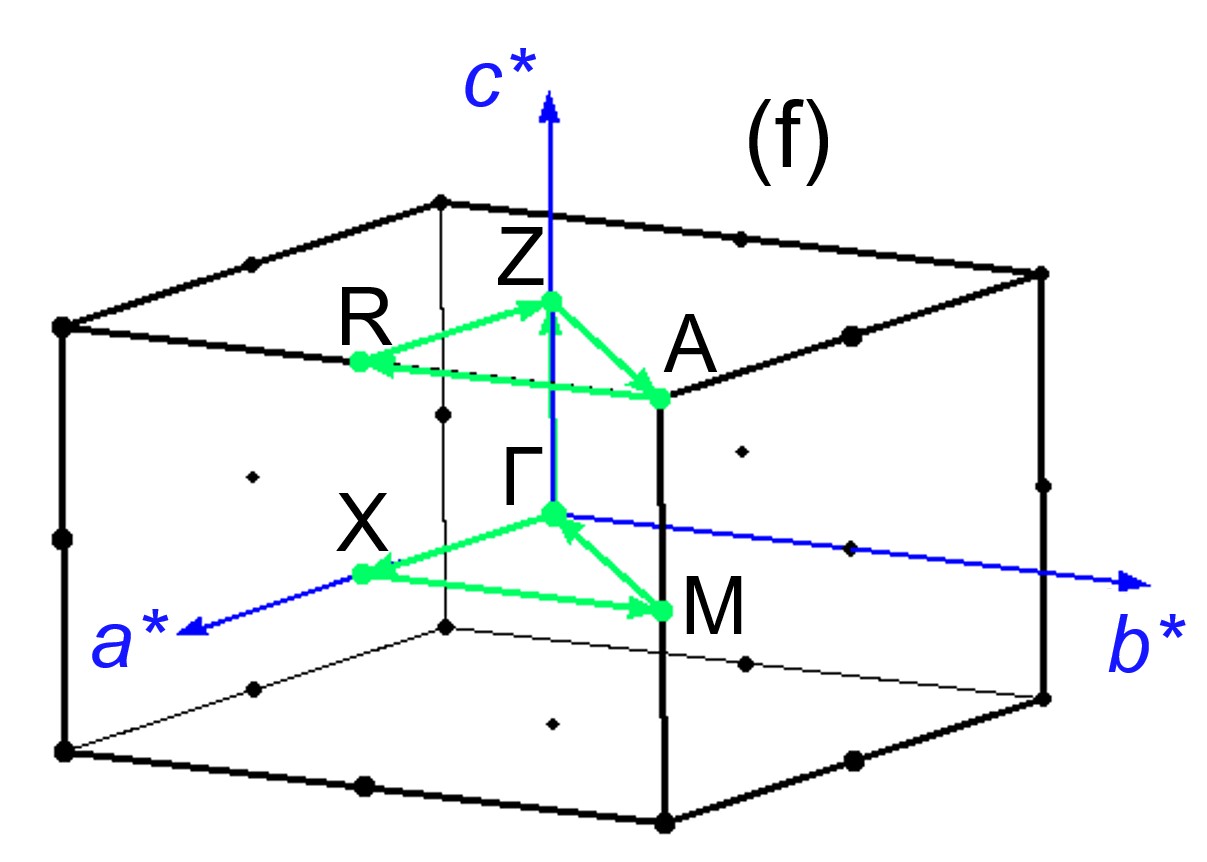}}
\caption{\label{figS5}Fat-band analysis for FeSe under (a) ambient pressure, (b) 4 GPa of hydrostatic pressure, (c) out-of-plane uniaxial compression, and (d) in-plane uniaxial compression calculated using maximally localized Wannier functions. Each graph is aligned to the Fermi energy, indicated by a dotted red line. A magnified view of the region around the Fermi level is shown in (e). The Brillouin zone used for calculation is shown in (f).}
\end{figure}

\begin{figure}[h]
\includegraphics[width=\textwidth]{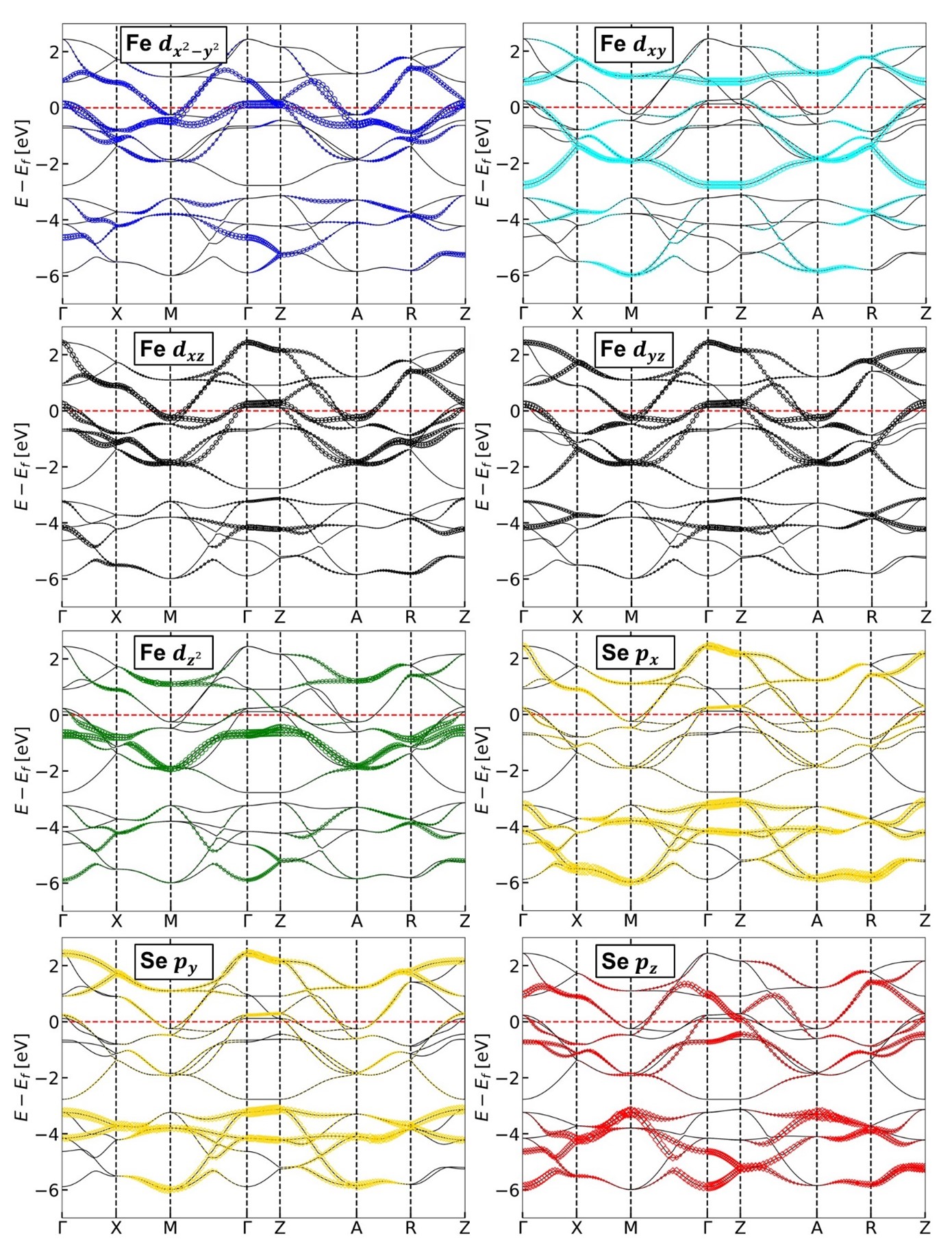}
\caption{\label{figS6}Fat-band analysis for FeSe under ambient pressure conditions.}
\end{figure}

\begin{figure}[h]
\includegraphics[width=\textwidth]{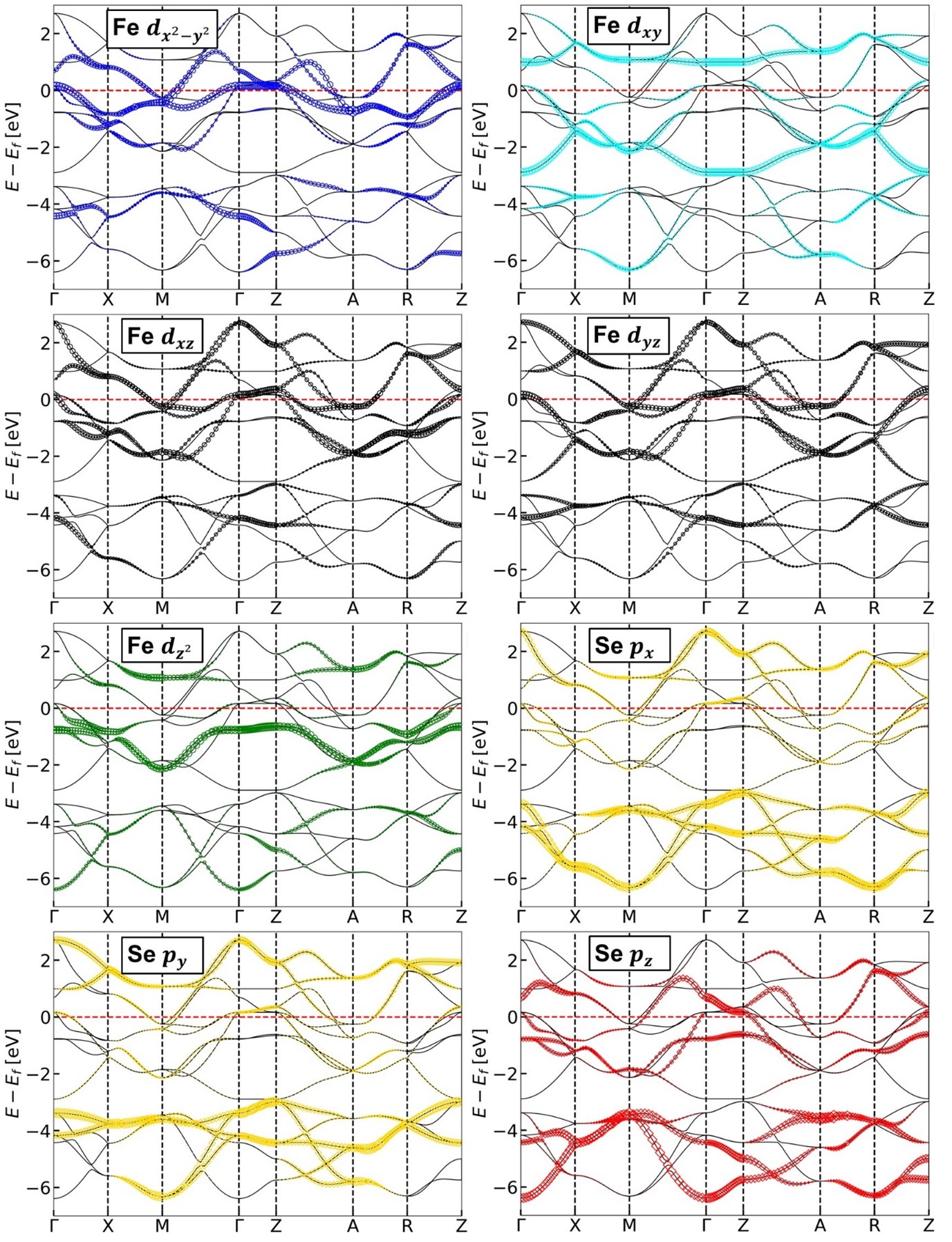}
\caption{\label{figS7}Fat-band analysis for FeSe under hydrostatic compression at 4 GPa.}
\end{figure}

\begin{figure}[h]
\includegraphics[width=\textwidth]{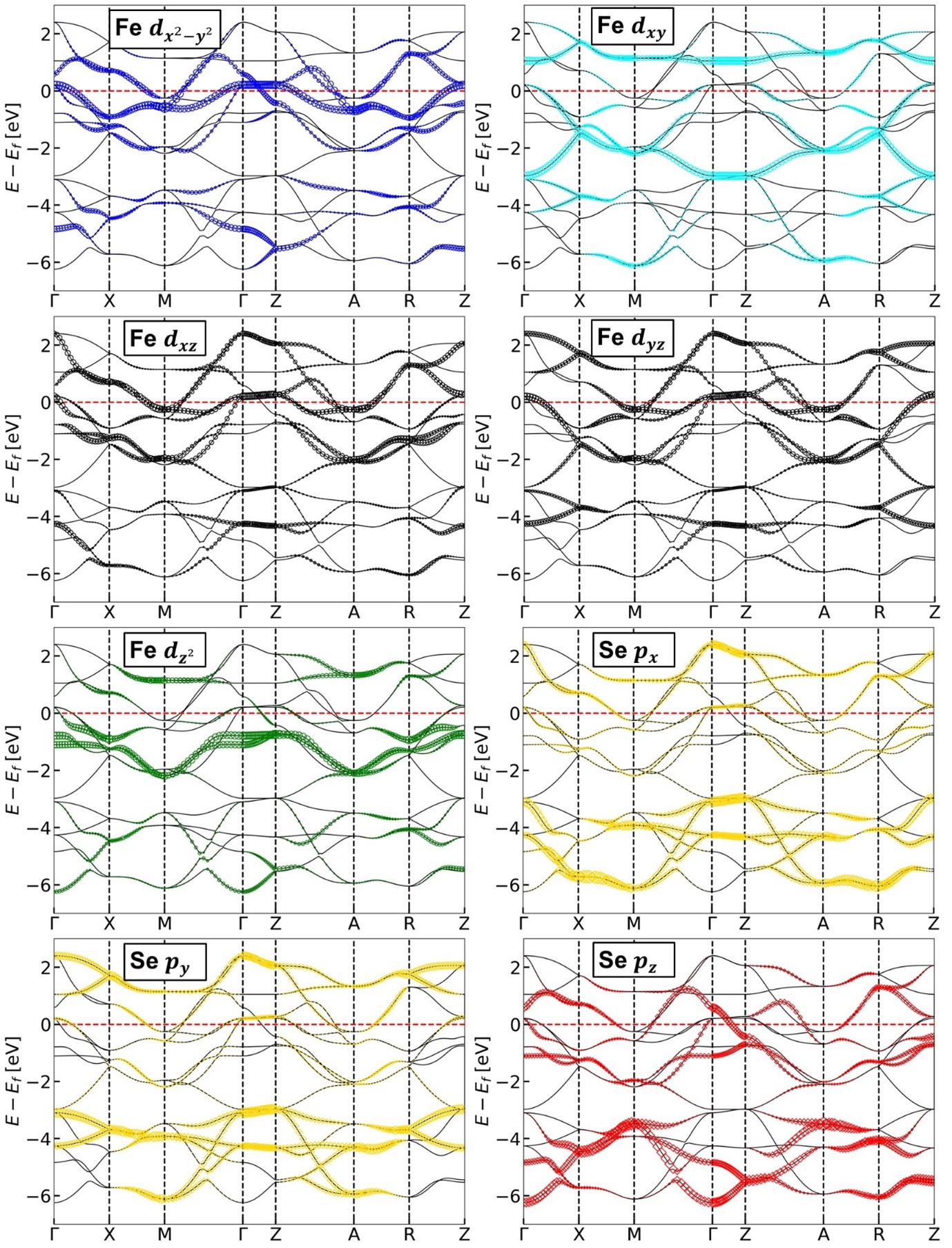}
\caption{\label{figS8}Fat-band analysis for FeSe under in-plane uniaxial compression at 4 GPa. We see a dispersive band near $E_F$ in the $\Gamma$-Z path, to which the Fe $d_{x^2-y^2}$ and Se $p_z$ orbitals contribute.}
\end{figure}

\begin{figure}[h]
\includegraphics[width=\textwidth]{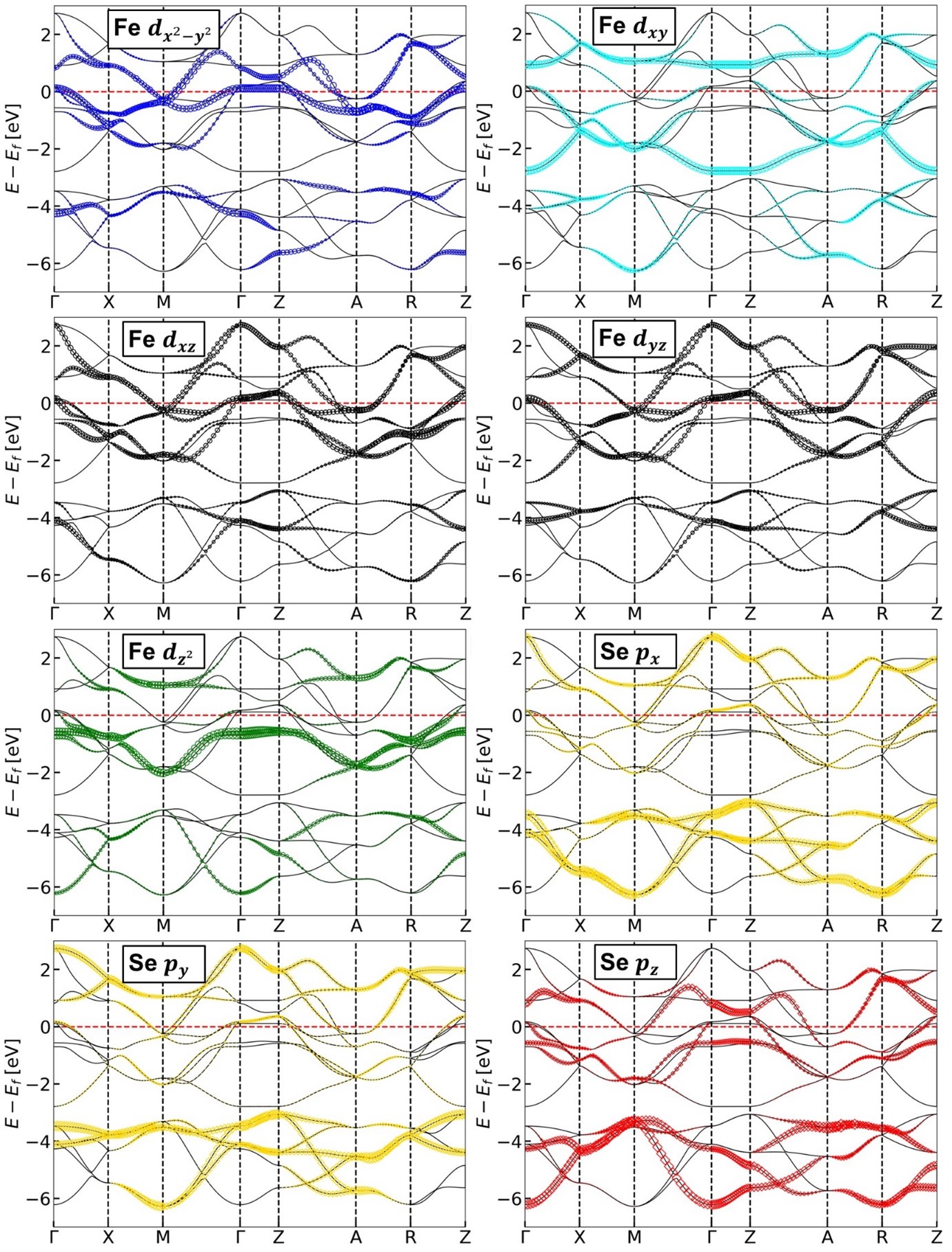}
\caption{\label{figS9}Fat-band analysis for FeSe under out-of-plane uniaxial compression at 4 GPa.}
\end{figure}

\end{document}